\documentclass[sn-nature]{sn-jnl}

\usepackage{graphicx}%
\usepackage{multirow}%
\usepackage{amsmath,amssymb,amsfonts}%
\usepackage{amsthm}%
\usepackage{mathrsfs}%
\usepackage[title]{appendix}%
\usepackage{xcolor}%
\usepackage{textcomp}%
\usepackage{manyfoot}%
\usepackage{booktabs}%
\usepackage{algorithm}%
\usepackage{algorithmicx}%
\usepackage{algpseudocode}%
\usepackage{listings}%

\usepackage{adjustbox}
\usepackage{xcolor}
\usepackage{longtable}
\usepackage{tabularx}
\usepackage{rotating} 
\usepackage{soul} 
\usepackage{makecell} 
\usepackage[algo2e]{algorithm2e} 
\usepackage{amsfonts}
\usepackage{gensymb}
\usepackage{float}
\restylefloat{table}

\theoremstyle{thmstyleone}%
\theoremstyle{thmstyletwo}%

\theoremstyle{thmstylethree}%

\begin{document}

\title{Stochastic Analysis of Cybersecurity Defense Strategies Under Single Attack Scenario}


\author[1]{\fnm{Song-Kyoo} \sur{Kim}} \email{amang@mpu.edu.mo}

\affil[1]{\orgdiv{Faculty of Applied Sciences}, \orgname{Macao Polytechnic University}, \orgaddress{\street{R. de Luis Gonzaga Gomes}, \city{Macau}, \country{SAR}}}


\abstract{
This research presents a novel stochastic framework for proactive cybersecurity defense timing under a single attack scenario. The approach models the defense process as a continuous observation mechanism in which the defense instant and the subsequent observation slot follow independent exponential distributions. Laplace-Carson transforms combined with first-excess theory yield the joint detection function that brackets the attack moment. Marginalization under Markovian Poisson arrivals then produces the probability density of the defense moment and conditional expectations of pre-attack and post-attack observation times. These closed-form results enable quantitative assessment of defense timing sensitivity to threat intensity and support precise calibration of observation parameters for low-latency proactive measures. Major contributions include the explicit derivation of marginal distributions and expected values, visualization of defense moment density, and the bridging of stochastic duel methodology with practical cybersecurity applications. 
}

\keywords{stochastic defense strategies; cybersecurity; continuous observation process; first-excess theory; defense timing sensitivity; network security.} 



\maketitle
\section{Introduction}
Cyberwarfare frequently arises from wars and political conflicts between nations \cite{LT01, LT08}. A possible correlation could link the frequency of a mention for an attacker in scientific literature to its actual incident rate. Furthermore, cyberattacks commonly occur on politically significant dates \cite{LT05}. Identification of suitable features from unstructured big data forms a crucial component within the proposed framework. Rapid advancement of Internet technology together with accelerating digital transformation has rendered cyberspace a vital element of modern society. Yet this digitization simultaneously escalates security threats, as various types of cyberattacks pose significant risks to personal privacy, business operations, and even national infrastructure with unprecedented frequency and complexity \cite{RE01, C005}. Achieving enhanced results in comprehending cyberattack time series relationships, a finding commonly reported in existing research, necessitates specialized preprocessing or the extraction of useful features from input data sources. Moreover, studies on cyberthreat intelligence mining \cite{RE09,A004} have examined cybersecurity related entities and events, cyberattack tactics, techniques and procedures, hacker profiles, indicators of compromise, vulnerability exploits, malware implementation, and threat hunting, while also presenting a comprehensive review of the state of the art. Prediction of cyberattacks that inject malicious signals into physical components or communication networks relies primarily on the intersection of two ellipsoid sets, with emphasis on replay attacks and bias injection attacks \cite{RE10, A005}. 

In the high stakes arena of modern cybersecurity, where governments and enterprises face relentless digital sieges reminiscent of historical battles against invisible foes, the single attack scenario stands as a powerful conceptual anchor that transforms overwhelming complexity into actionable clarity. It envisions an assault not as an endless barrage of polymorphic intrusions but as one decisive, isolated strike—the critical instant when an attacker crosses the threshold from preparation to execution. Even amid potential waves of attempts, the framework deliberately isolates the first incursion, discarding all subsequent efforts to concentrate analytical power on that singular moment of vulnerability. This elegant reduction mirrors the sentinel’s vigil in ancient fortifications, where one well-timed alert could avert catastrophe, allowing the model to bracket the threat precisely between the moment of proactive defense and the immediate aftermath. By framing the attack as a one-time event of either triumph or failure, the approach illuminates the delicate timing of preliminary actions, empowering defenders to anticipate and intercept before damage escalates. This focused lens reveals profound insights into resource allocation and low-latency response strategies, ultimately bridging theoretical stochastic elegance with the urgent realities of safeguarding global digital infrastructure against the next unforeseen breach. 
The single attack scenario assumes that a cyberattack constitutes a one-time event of either success or failure, such that in cases of multiple attack attempts only the moment of the first attempt is retained as the attack time while the rest of the attempts are discarded. 

The paper presents a novel stochastic framework for analyzing cybersecurity defense strategies under a single attack scenario. It models the proactive defense process as a continuous observation mechanism wherein the moment of defense and the subsequent observation slot follow exponential distributions. By employing Laplace-Carson transforms together with first-excess theory, the joint characteristics of the defense observations that bracket the attack instant are derived explicitly. Marginalization under a Markovian Poisson attack arrival process yields the probability density function of the defense moment along with conditional expectations for both pre-attack and post-attack observation times. These analytical results, supported by visualization of the marginal density, enable quantitative assessment of defense timing sensitivity to varying attack rates. The framework contributes valuable closed-form insights that facilitate calibration of observation parameters for enhanced proactive cybersecurity measures, bridging theoretical stochastic analysis with practical needs for low-latency threat mitigation in contemporary network environments.

\section{Preliminaries} \label{sec02}
\begin{figure}[H]
\centering
\includegraphics[width=.9\columnwidth]{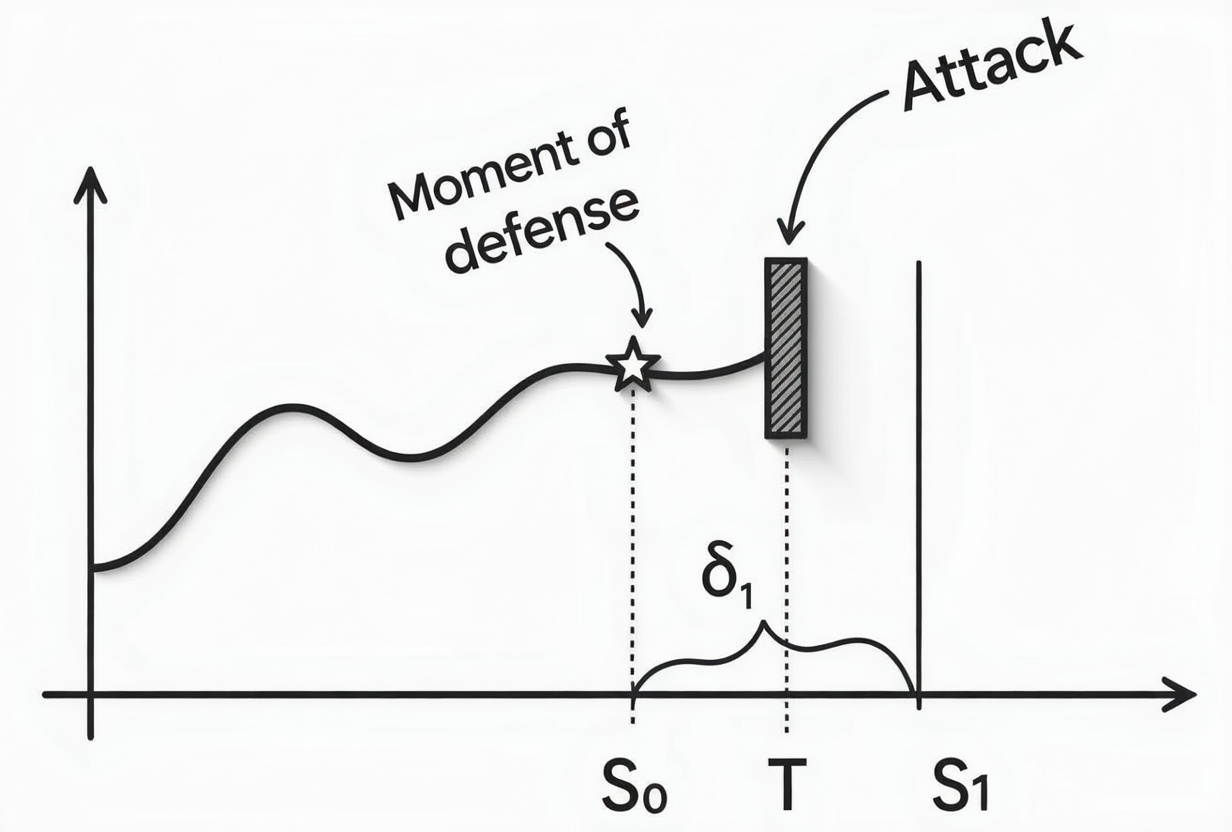}
 \caption{The observation process until a single attack event.}
 \label{Fig001}
\end{figure}

This section establishes the preliminaries essential for the stochastic analysis of cybersecurity defense strategies under a single attack scenario. The attack arrival time is modeled as a one-time event within a continuous observation framework that brackets the critical moment through sequential instants immediately before and after the attack. Alternatively, the duration between the first observation $S_0$, taken just before the attack, and the second observation $S_1$, taken right after the attack, represents the inter-arrival time. The moment $S_0$ is the moment of the defense when preliminary action for cybersecurity defense is taken. The duration between $S_0$ and $S_1$ represents the inter-arrival time $\delta_1$ following the first observation $S_0$ (see Fig.~\ref{Fig001}). The attack moment $T$ shall be positioned between these two observations (i.e., between $S_0$ and $S_1$). It is noted that the post-attack moment has not actually occurred when an attack is attempted. 
Fig.~\ref{Fig001} illustrates the continuous observation process under a single attack scenario, depicting the defense moment immediately preceding the attack and the subsequent observation slot that brackets the critical instant. 

This schematic clarifies the temporal relationship between preliminary defensive actions and post-attack verification within the stochastic framework. The analysis proceeds by assuming exponential distributions for both the first observation time and the observation slot duration to derive the necessary transformation functions and joint detection characteristics. 

\subsection{Modeling of sequential observations in the single-attack scenario}
The continuous-time observation process for a single termination event in the cybersecurity defense model assumes that the first observation moment and the subsequent time slot follow independent exponential distributions, thereby enabling explicit derivation of the joint function that characterizes the pair of observations bracketing the attack instant. Hence, the moment of the defense follows an exponential distribution with the rate $\mu_0$ and the accumulative probability is defined as follows:

\begin{equation}
\mathbb{P} \left[S_0 \leq t_0 \right]
=1-e^{-\mu_0 t_0}, t_0 \geq 0,
\end{equation}

\noindent and the time slot for determining the correct prediction after the moment of the defense $\delta_1$ also follows an exponential distribution with the parameter $\mu_1$ which could be defined as follows:

\begin{equation}
\mathbb{P} \left[\delta_1 \leq t_1 \right]
=1-e^{-\mu_1 t_1}, t_1 > 0,
\end{equation}

\noindent and the moment of the post-attack $S_1$ could be calculated based on $S_0$ and $\delta_1$ which is as follows:

\begin{equation}
S_1 = S_0+\delta_1, S_1 < S_0. 
\end{equation}

The Laplace transforms for the process for the pre-attack (first) observation $S_0$ and for the prediction time slot $\delta_1$ are constructed as follows:

\begin{equation}
\gamma_0\left(\theta\right)
=\mathbb{E}\left[e^{-\theta S_0}\right]
=\frac{\mu_0}{\mu_0+\theta},
\gamma_1\left(\theta\right)
=\mathbb{E}\left[e^{-\theta\delta_1}\right]
=\frac{\mu_1}{\mu_1+\theta}.
\label{eq004a}
\end{equation}


The join function of cyberattack prediction which provides the characteristics of the defense strategies (i.e., pre- and post-attack observations) could be defined as follows:

\begin{equation}
\varPhi_{p^\ast}\left(\theta_0,\theta_1\right)=\mathbb{E}\left[e^{-%
\theta_0S_0}e^{-\theta_1S_1}\cdot\boldsymbol{1}
_{\left\{S_0<p^\ast \leq S_1\right\}}\right],
\text{Re}\left(\theta_0\right)>0, \text{Re}\left(\theta_1\right)>0,
\label{eq003a}
\end{equation}

\noindent where $p^\ast$ is the moment of the attack where $T=p^\ast$. The system represent the the status of before and after the attack is happened. The univariate Laplace-Carson transform is applied as follows:

\begin{equation}
\widehat{\mathcal{L}}_p\left(\bullet\right)\left(u\right)=u\int^%
\infty_{p=0}e^{-up}\left(\bullet\right)dp, \text{Re}\left(u\right)>0, 
\label{eq004}
\end{equation}

\noindent with the inverse

\begin{equation}
\widehat{\mathcal{L}}_u^{-1}\left(\bullet\right)\left(p\right)=%
\mathcal{L}^{-1}\left(\bullet\frac{1}{u}\right), 
\label{eq006c}
\end{equation}

\noindent and

\begin{equation}
\widehat{\mathcal{L}}_u^{-1}\left(\bullet\right)\left(r\right)=\widehat{%
\mathcal{L}}_u^{-1}\left(\bullet\right)\left(p\right)\Big|_{p\rightarrow
r},
\label{eq007b}
\end{equation}

\noindent where $\mathcal{L}^{-1}$ is the inverse of the Laplace transform  \cite{BGG07, AMG002}. From (\ref{eq006c})-(\ref{eq007b}), these are supply the fundamental operational tools for mapping the joint function of cyberattack detection between the time domain and the transform domain. This pair of operators allows algebraic simplification of the complex stochastic expression and subsequent recovery of the original time-domain probability characteristics through standard inversion techniques.
The join function of cyberattack prediction $\varPhi_{p^\ast}\left(\theta_0,\theta_1\right)$ which characterizes the cybersecurity defense strategy could be constructed as follows:

\begin{equation}
\varPhi_{p^\ast}\left(\theta_0,\theta_1\right)
=\widehat{\mathcal{L}}_u^{-1}\left\{\varPsi_u\left(\theta_0,\theta_1\right)\right\}\left(p^\ast\right),
\label{eq006b}
\end{equation}

\noindent where

\begin{equation}
\varPsi_u\left(\theta_0,\theta_1\right)
=\widehat{\mathcal{L}}_p\left\{%
\varPhi_{p^\ast}\left(\theta_0,\theta_1\right)\right\}\left(u\right),
\text{Re}\left(\theta_0\right)>0,
\text{Re}\left(\theta_1\right)>0.
\label{eq010a}
\end{equation}

\noindent which is the transformed functional of the joint function from (\ref{eq006b}).

\ 

\noindent \textbf{Theorem:} From (\ref{eq006b})-(\ref{eq010a}), we have:
\begin{equation}
\varPhi_{p^\ast}\left(\theta_0,\theta_1\right)
=\widehat{\mathcal{L}}_u^{-1}\left\{\varGamma_0\cdot\gamma_1-\varGamma_0\cdot\varGamma_1
\right\}\left(p^\ast\right),
\label{eq006}
\end{equation}

\noindent where
\begin{equation}
\varGamma_0 =\frac{\mu_0}{u+\mu_0+\theta_0+\theta_1},
\gamma_1 = \frac{\mu_1}{\mu_1+\theta_1},
\varGamma_1 = \frac{\mu_1}{u+\mu_1+\theta_1}.
\label{eq007}
\end{equation}

\begin{proof}
To prove the formula (\ref{eq006}), we first notice that

\begin{equation}
\boldsymbol{1}_{\left\{\nu^\ast\left(p\right)=1\right\}}=\left(%
\boldsymbol{1}_{\left\{S_0 < p\right\}}\boldsymbol{1}_{\left\{S_1\geq p\right\}}\right)
\label{eq011}
\end{equation}

\noindent could be transformed by iterating the integral of (\ref{eq006}). Then we have:

\begin{equation}
\widehat{\mathcal{L}}_{pq}\left\{\boldsymbol{1}_{\left\{\nu^\ast\left(p%
\right)=1\right\}}\right\}\left(u\right)=e^{-{uS_0}} - e^{-{uS_1}},
\label{eq012a}
\end{equation}

\noindent by the first exceed theory for the continuous random variables \cite{BGG07, AMG001}. From (\ref{eq004}) and (\ref{eq012a}), the functional from (\ref{eq010a}) could be calculated as follows:

\begin{equation*}
\varPsi_u\left(\theta_0,\theta_1\right)
=\mathbb{E}\left[e^{-\theta_0S_0}e^{-\theta_1S_1}\cdot%
\left(e^{-uS_0}-e^{-uS_1}\right)\right]
=L_1 - L_2,
\end{equation*}

\begin{equation*}
=\mathbb{E}\left[
e^{-\left(\theta_0+\theta_1+u\right)S_0}
e^{-\delta_1} \right] 
-\mathbb{E}\left[ e^{-\left(\theta_0+\theta_1+u\right)S_0}
e^{-\left(\theta_1+u\right)\delta_1} \right],
\end{equation*}

\noindent and
\begin{equation*}
L_1=\mathbb{E}\left[
e^{-\left(\theta_0+\theta_1+u\right)S_0}
e^{- \theta_1\delta_1} \right] = \varGamma_0\cdot\gamma_1
\end{equation*}


\begin{equation*}
L_2=\mathbb{E}\left[ e^{-\left(\theta_0+\theta_1+u\right)S_0}
e^{-\left(\theta_1+u\right)\delta_1} \right]
=\varGamma_0\cdot\varGamma_1, 
\end{equation*}

\noindent where
\begin{equation}
\varGamma_0=\gamma_0\left(\theta_0+\theta_1+u\right),\gamma_1=\gamma_1\left(%
\theta_1\right),\varGamma_1=\gamma_1\left(\theta_1+u\right).
\label{eq014a}
\end{equation}

From (\ref{eq004a}), the formulas from (\ref{eq014a}) could be reconstructed as follows:

\begin{equation*}
\varGamma_0 =\frac{\mu_0}{u+\mu_0+\theta_0+\theta_1},
\gamma_1 = \frac{\mu_1}{\mu_1+\theta_1},
\varGamma_1 = \frac{\mu_1}{u+\mu_1+\theta_1}.
\end{equation*}
\end{proof}

Since the probability distributions for both the defense moment $S_0$ and the post-attack moment $S_1$ possess Markovian properties, the joint function of cyberattack detection can be expressed explicitly. 
Recall from (\ref{eq006}), we have:

\begin{equation*}
\varPhi_{p^\ast}\left(\theta_0,\theta_1\right)
=\widehat{\mathcal{L}}_u^{-1}\left\{\varGamma_0\cdot\gamma_1\right\}%
-\widehat{\mathcal{L}}_u^{-1}\left\{ \varGamma_0\cdot\varGamma_1\right\},%
\end{equation*}

\noindent and

\ \ \ \ \ \ \ \ \ \ \ \ \ \ \
$\widehat{\mathcal{L}}_u^{-1}\left\{\varGamma_0\cdot\gamma_1\right\}=
\frac{\mu_0\mu_1}{\left(\mu_1+\theta_1\right)\left(\mu_0+\theta_0+\theta_1%
\right)}-\frac{\mu_0\mu_1e^{-\left(\mu_0+\theta_0+\theta_1\right)p}}{\left(%
\mu_1+\theta_1\right)\left(\mu_0+\theta_0+\theta_1\right)}$,

\ 

\noindent and 

\ \ \ \ \ \ \ \ \ \ \ \ \ \ \
$\widehat{\mathcal{L}}_u^{-1}\left\{\varGamma_0\cdot\varGamma_1\right\}=\frac{\mu_0\mu_1}{\left(\mu_1+\theta_1\right)\left(\mu_0+\theta_0+\theta_1%
\right)}-\frac{\mu_0\mu_1e^{-\left(\mu_1+\theta_1\right)p}}{\left(\mu_1+%
\theta_1\right)\left(\mu_0-\mu_1+\theta_0\right)}$

\ 

\ \ \ \ \ \ \ \ \ \ \ \ \ \ \ \ \ \ \ \ \ \ \ \ \ \ \ \ \ \ \ \ \ \ \ \ \ \ \ \ \ \ \ \ \ \ \ \ \ \ \ \ \ \ \ \ \ \ \ \ \ \ \ \ \ \ \ \ \ \ \ \ \ \ 
$ +\frac{\mu_0\mu_1e^{-\left(%
\mu_0+\theta_0+\theta_1\right)p}}{\left(\mu_0+\theta_0+\theta_1\right)\left(%
\mu_0-\mu_1+\theta_0\right)}.$

\

Therefore, the cyberattack prediction joint function $\varPhi_p\left(\theta_0,\theta_1\right)$ could be constructed as follows:
\begin{equation}
\varPhi_p\left(\theta_0,\theta_1\right)=\frac{\mu_0\mu_1e^{-\left(\mu_1+%
\theta_1\right)p}}{\left(\mu_1+\theta_1\right)\left(\mu_0-\mu_1+\theta_0%
\right)}
-\left(\frac{\mu_0\mu_1e^{-\left(\mu_0+\theta_0+\theta_1\right)p}}{%
\mu_0+\theta_0+\theta_1}\right)\left(\frac{1}{\mu_1+\theta_1}+\frac{1}{\mu_0-%
\mu_1+\theta_0}\right),
\label{eq013}
\end{equation}

\noindent and the probability density function (pdf) for the defense moment $g\left(s\right)$ with the attack moment $p^\ast$ (i.e., $g_{p^\ast}(s)$) could be found by Inverse Laplace Transform (or Bromwich integral) which is as follows :

\begin{equation}
g\left(s,p\right)=\frac{\partial}{\partial s}\left\{\mathbb{E}\left[\boldsymbol{1}_{\left\{S_0<s \right\}}\cdot\boldsymbol{1}_{\left\{S_0< p \leq
S_1\right\}}\right]\right\}=\mathcal{L}_{\theta_0}^{-1}\left\{\varPhi_{p}\left(\theta_0,0\right)\right\}, 
\label{eq016a}
\end{equation}

\noindent and the cumulative distribution function (CDF) for the moment of defense $G\left(s,p\right)$ could be found as follows:

\begin{equation}
G(s,p)=\mathcal{L}_{\theta_0}^{-1}\left\{\frac{\varPhi_{p}\left(\theta_0,0\right)}{\theta_0}\right\}
=\widehat{\mathcal{L}}_{\theta_0}^{-1}\left\{\varPhi_{p}\left(\theta_0,0\right)\right\},
\text{Re}(\theta_0) > 0.
\label{eq017}
\end{equation}

It is noted that $g_{p^\ast}\left(s\right)$ has been has been rewritten as $g\left(s,p\right)$ for the mathematical convenient. From (\ref{eq016a}), the mean of the first observation moment (i.e., the moment of the defense) could be calculated as follows:

\begin{equation*}
\mathbb{E}\left[S_0\cdot\boldsymbol{1}_{\left\{S_0<p^\ast\leq
S_1\right\}}\right]=\int^\infty_{s=0}s\cdot
g_{p^\ast}(s)ds.
\end{equation*}

Alternatively, the mean of the first observation moment directly calculated by taking the derivative of the joint function:

\begin{equation}
\mathbb{E}\left[S_0\cdot\boldsymbol{1}_{\left\{S_0<p^\ast\leq
S_1\right\}}\right]
=-\frac{\partial\varPhi_{p^\ast}\left(\theta_0,0\right)}{\partial\theta_0}
\bigg|_{\theta_0=0},
\label{eq017a}
\end{equation}

\noindent and, similarly, the mean of the post-attack moment $S_1$ could be calculated as follows:

\begin{equation*}
\mathbb{E}\left[S_1\cdot\boldsymbol{1}_{\left\{S_0<p^\ast\leq
S_1\right\}}\right]
=-\frac{\partial\varPhi_{p^\ast}\left(0,\theta_1\right)}{\partial\theta_1}
\bigg|_{\theta_1=0}.
\end{equation*}

The preliminaries in this section provide the basics for the stochastic cybersecurity defense model under a single attack scenario. It models the defense process using sequential observations bracketing the attack moment, where both the defense instant and the following slot are exponentially distributed. The joint function of these observations is explicitly derived via Laplace-Carson transforms.

\section{Markovian modeling of defense strategies} \label{sec03}
The preliminaries have established a rigorous continuous-time framework for modeling the observation process under a single attack scenario, deriving the joint function that fully characterizes the defense instants bracketing the attack moment. Building upon these foundations, this section investigates Markovian attack behavior. By exploiting the memoryless property of the stochastic process, the analysis incorporates a Poisson arrival process for cyberattacks. This Markovian perspective enables straightforward marginalization of the joint function, yielding explicit marginal probability density functions for the defense moment together with conditional expectations of the pre-attack and post-attack observation times.

\subsection{Analysis of the defense moment}
This section addresses the prediction of the moment of defense, identified as the first observation taken just before a cyberattack in the single attack scenario. By considering the joint transform in the limiting case for the post-attack parameter and averaging over the Poisson attack arrival process, the marginal behavior is obtained. Inversion techniques yield the probability density function of the defense instant, which is visualized to reveal its variation with attack intensity. Differentiation of the marginal expectation further delivers the expected value of the defense moment. These results supply essential quantitative tools for calibrating defense observation parameters and improving the timeliness of proactive cybersecurity measures. The moment of the pre-attack (i.e., the moment of defense) could be analyzed from the joint function when $\theta_1 \rightarrow 0$. 
From (\ref{eq013}), the moment of defense can be determined as follows (i.e., $\varPhi\left(\theta_0,0 ; p \right):=\varPhi_{p}\left(\theta_0,0\right)$): 

\begin{equation}
\varPhi\left(\theta_0,0 ; p \right)=\frac{\mu_0\mu_1e^{-\mu_1p}}{\mu_1\left(%
\mu_0-\mu_1+\theta_0\right)}-\left(\frac{\mu_0\mu_1e^{-\left(\mu_0+\theta_0%
\right)p}}{\mu_0+\theta_0}\right)\left(\frac{1}{\mu_1}+\frac{1}{\mu_0-\mu_1+%
\theta_0}\right),
\label{eq014}
\end{equation}

\noindent and the joint function under the marginal single-attack scenario with the random attack arrival time $T$ is analyzed by applying double expectations:

\begin{equation*}
\mathbb{E}\left[\varPhi\left(\theta_0,0;T\right)\right]=\mathbb{E}\left[%
\mathbb{E}\left[\varPhi\left(\theta_0,0;T\right)\right]\right],
\end{equation*}

\noindent then we have: 

\begin{equation}
\mathbb{E}\left[\varPhi\left(\theta_0,0;T\right)\right]=\int^\infty_{p=0}%
\varPhi\left(\theta_0,0;p\right)dF(p).
\label{eq015}
\end{equation}

Since the attack arrival time $  T  $ follows a Markovian Poisson process with rate $\lambda$, the Laplace transform of the random variable $T$ is as follows:

\begin{equation}
\mathbb{E}\left[e^{-\theta T}\right]=\int^\infty_{p=0}\left\{e^{-\theta
p}\right\}dF\left(p\right)=\frac{\lambda}{\lambda+\theta}, \text{Re}\left( \theta \right) \geq 0.
\label{eq016}
\end{equation}

From (\ref{eq015}) and (\ref{eq016}), the joint function under the marginal single attack scenario could be found as follows:

\ 

\ \ \ \ 
$\mathbb{E}\left[\varPhi\left(\theta_0,0;T\right)\right]=\int^\infty_{p=0}%
\left\{\frac{\mu_0e^{-\mu_1p}}{\mu_0-\mu_1+\theta_0}-\left(\frac{1}{\mu_1}+%
\frac{1}{\mu_0-\mu_1+\theta_0}\right)\frac{\mu_0\mu_1e^{-\left(\mu_0+\theta_0%
\right)p}}{\mu_0+\theta_0}\right\}dF(p)$

\ 

\ \ \ \ \ \ \ \ \ \ \ \ \ \ \ \ \ \ \ \ \ \ \ 
$=\left(\frac{\mu_0}{\mu_0-\mu_1+\theta_0}\right)\int^\infty_{p=0}\left\{e^{-\mu_1p}\right\}dF(p)$

\

\ \ \ \ \ \ \ \ \ \ \ \ \ \ \ \ \ \ \ \ \ \ \ \ \ \ \ \ \ \ \ \ \ \
$-\left(\frac{\mu_0\mu_1}{\mu_1\left(\mu_0+\theta_0\right)}+\frac{\mu_0%
\mu_1}{\left(\mu_0+\theta_0\right)\left(\mu_0-\mu_1+\theta_0\right)}\right)%
\int^\infty_{p=0}\left\{e^{-\left(\mu_0+\theta_0\right)p}\right\}dF(p)$

\ 

\ \ \ \ \ \ \ \ \ \ \ \ \ \ \ \ \ \ \ \ \ \ \ 
$=\frac{\mu_0\lambda}{\left(\mu_0-\mu_1+\theta_0\right)\left(\lambda+\mu_1\right)}-\left\{\frac{\mu_0}{\mu_0+\theta_0}-\frac{\mu_0}{\lambda+\mu_0+\theta_0}\right\}$

\

\ \ \ \ \ \ \ \ \ \ \ \ \ \ \ \ \ \ \ \ \ \ \ \ \ \ \ \ \ \ \ \ \ \
$-\left\{-\frac{\mu_0}{\left(\mu_0+\theta_0\right)}+\frac{\mu_0\mu_1}{\left(%
\mu_1+\lambda\right)\left(\lambda+\mu_0+\theta_0\right)}+\frac{\mu_0\lambda}{%
\left(\mu_1+\lambda\right)\left(\mu_0-\mu_1+\theta_0\right)}\right\},$

\ 

\ \ \ \ \ \ \ \ \ \ \ \ \ \ \ \ \ \ \ \ \ \ \ 
$=\frac{\mu_0\lambda}{\left(\mu_0-\mu_1+\theta_0\right)\left(\lambda+\mu_1%
\right)}-\frac{\mu_0}{\mu_0+\theta_0}+\frac{\mu_0}{\lambda+\mu_0+\theta_0}$

\

\ \ \ \ \ \ \ \ \ \ \ \ \ \ \ \ \ \ \ \ \ \ \ \ \ \ \ \ \ \ \ \ \ \
$+\frac{\mu_0}{\left(\mu_0+\theta_0\right)}-\frac{\mu_0\mu_1}{\left(\mu_1+%
\lambda\right)\left(\lambda+\mu_0+\theta_0\right)}-\frac{\mu_0\lambda}{\left(%
\mu_1+\lambda\right)\left(\mu_0-\mu_1+\theta_0\right)},$

\noindent then we finally conclude:

\begin{equation}
\mathbb{E}\left[\varPhi\left(\theta_0,0;T\right)\right]=\frac{\mu_0\lambda}{%
\left(\mu_1+\lambda\right)\left(\lambda+\mu_0+\theta_0\right)},  \text{Re}\left(\theta_0\right)>0.
\label{eq017}
\end{equation}

\noindent then the marginal pdf of the moment of defense $g(s,p)$ from (\ref{eq016a}) with respect to the Markovian attack moment $T$ could be calculated as follows:

\begin{equation}
\mathbb{E}\left[g(s,T)\right]
=\mathcal{L}_{\theta_0}^{-1}\left\{
\mathbb{E}\left[\varPhi\left(\theta_0,0;T\right)\right]
\right\} 
=\frac{\mu_0\lambda\cdot
e^{-\left(\mu_0+\lambda\right)s}}{\left(\mu_1+\lambda\right)\left(\mu_0+%
\lambda\right)}.
\label{eq022}
\end{equation}


\

\begin{figure}[H]
\centering
\includegraphics[width=.9\columnwidth]{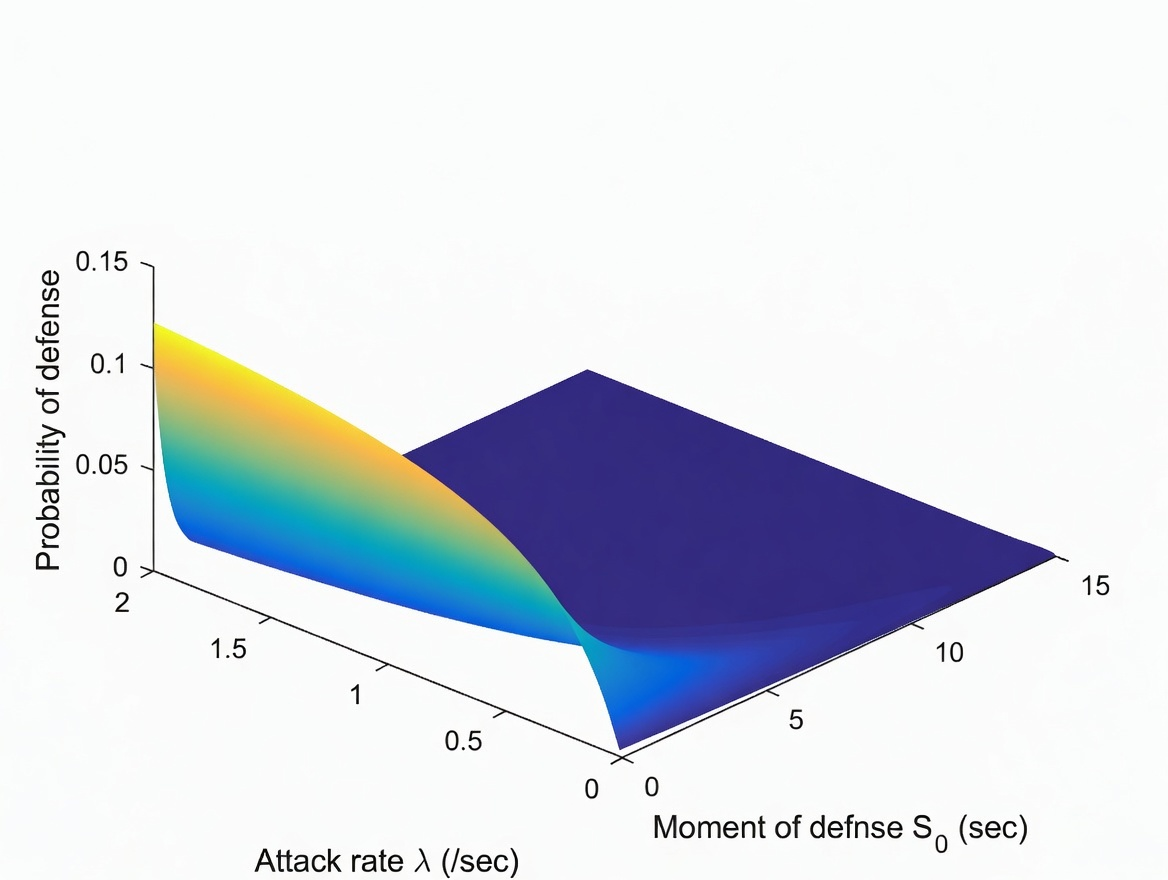}
 \caption{Marginal probability density function for the moment of defense $S_0$.}
 \label{Fig002}
\end{figure}

Additionally, the marginal pdf $\mathbb{E}\left[g(s,T)\right]$ with the respect of the attack rate $\lambda$ could be visualized as shown in Fig. \ref{Fig002}. This figure portrays the marginal probability density function for the moment of defense $S_0$ within the stochastic analysis of cybersecurity defense strategies. This visualization demonstrates the density surface across the attack rate $\lambda$ and defense moment $S_0$, highlighting how the probability distribution of the preliminary defense action evolves with varying threat intensities. 

\subsection{Analysis of the post-attack moment}

Similarly, the moment of the post-attack $S_1$ could be found when $\theta_0 \rightarrow 0$. Finding the moment of the post-attack could be started from (\ref{eq013}) which is:

\begin{equation}
\varPhi_p\left(0,\theta_1\right)=\frac{\mu_0\mu_1e^{-\left(\mu_1+\theta_1%
\right)p}}{\left(\mu_1+\theta_1\right)\left(\mu_0-\mu_1\right)}-\left(\frac{%
\mu_0\mu_1e^{-\left(\mu_0+\theta_1\right)p}}{\mu_0+\theta_1}\right)\left(%
\frac{1}{\mu_1+\theta_1}+\frac{1}{\mu_0-\mu_1}\right).
\label{eq018}
\end{equation}

From (\ref{eq015}) and (\ref{eq018}), the marginal probability of the the post-attack could be constructed as follows:

\ 

\ \ \ \ 
$\mathbb{E}\left[\varPhi\left(0,\theta_1;T\right)\right]=\int^\infty_{p=0}\left\{\frac{\mu_0\mu_1e^{-\left(\mu_1+\theta_1\right)p}}{\left(\mu_1+%
\theta_1\right)\left(\mu_0-\mu_1\right)}-\left(\frac{1}{\mu_1+\theta_1}+%
\frac{1}{\mu_0-\mu_1}\right)\frac{\mu_0\mu_1e^{-\left(\mu_0+\theta_1%
\right)p}}{\mu_0+\theta_1}\right\}dF(p)$

\ 

\ \ \ \ \ \ \ \ \ \ \ \ \ \ \ \ \ \ \ \ \ \ \ 
$=\frac{\mu_0\mu_1\lambda}{\left(\mu_0-\mu_1\right)\left(\mu_1+\theta_1%
\right)\left(\lambda+\mu_1+\theta_1\right)}$

\ \ \ \ \ \ \ \ \ \ \ \ \ \ \ \ \ \ \ \ \ \ \ \ \ \ \ \ \ \ \ \ \ \ \ \ \ \ \ \ \ \ \ \ \ \ \ \ \ 
$-\frac{\mu_0\mu_1\lambda}{\left(\mu_1 +\theta_1\right)\left(\mu_0+\theta_1\right)\left(\lambda+\mu_0+\theta_1\right)}$

\ \ \ \ \ \ \ \ \ \ \ \ \ \ \ \ \ \ \ \ \ \ \ \ \ \ \ \ \ \ \ \ \ \ \ \ \ \ \ \ \ \ \ \ \ \ \ \ \ \ \ \ \ \ \ \ \ \ \ \ \ \ \ \ \ \ \ \ \ \ \ \ \ \ 
$-\frac{\mu_0\mu_1\lambda}{\left(\mu_0-\mu_1\right)\left(\mu_0+%
\theta_1\right)\left(\lambda+\mu_0+\theta_1\right)}$

\ 

\ \ \ \ \ \ \ \ \ \ \ \ \ \ \ \ \ \ \ \ \ \ \ 
$=\frac{\mu_0\mu_1\lambda}{\left(\mu_0-\mu_1\right)\left(\lambda+\mu_0-\mu_1%
\right)\left(\mu_1+\theta_1\right)}-\frac{\mu_0\mu_1}{\left(\mu_0-\mu_1%
\right)\left(\mu_0+\theta_1\right)}+\frac{\mu_0\mu_1}{\left(\mu_0-\mu_1+%
\lambda\right)\left(\mu_0+\lambda+\theta_1\right)}$

\ 

\ \ \ \ \ \ \ \ \ \ \ \ \ \ \ \ \ \ \ \ \ \ \ 
$=\frac{\mu_0\mu_1}{\left(\mu_0-\mu_1\right)\left(\mu_1+\theta_1\right)}-\frac{\mu_0\mu_1\lambda}{\left(\mu_0-\mu_1\right)\left(\lambda+\mu_0-\mu_1\right)\left(\mu_1+\theta_1\right)}-\frac{\mu_0\mu_1}{\left(\mu_0-\mu_1+\lambda\right)\left(\mu_0+\lambda+\theta_1\right)}$

\ 

Therefore, 
\begin{equation}
\label{eq019}
\end{equation}

\ \ \ \ \ \ \ \ \ \ 
$\mathbb{E}\left[\varPhi\left(0,\theta_1;T\right)\right]
=\frac{\mu_0\mu_1}{\left(\mu_0-\mu_1\right)\left(\mu_1+\theta_1\right)}$

\ 

\ \ \ \ \ \ \ \ \ \ \ \ \ \ \ \ \ \ \ \ \ \ \ \ \ \ \ \ \ \ \ \ \ \ \ \ \ \ 
$-\frac{\mu_0\mu_1\lambda}{\left(\mu_0-\mu_1\right)\left(\lambda+\mu_0-\mu_1\right)\left(\mu_1+\theta_1\right)}-\frac{\mu_0\mu_1}{\left(\mu_0-\mu_1+\lambda\right)\left(\mu_0+\lambda+\theta_1\right)}$.

\ 

The marginal mean for the moment of the post-attack under the single attack scenario could be determined as follows:

\begin{equation}
\mathbb{E}\left[S_1\cdot\boldsymbol{1}_{\left\{S_0<T\leq
S_1\right\}}\right]=-\frac{d\mathbb{E}\left[\varPhi\left(0,\theta_1;T\right)%
\right]}{d\theta_1} \bigg|_{\theta_1=0},
\label{eq020}
\end{equation}

\noindent and the derivative of (\ref{eq020}) could be found as follows:

\ 

\ \ \ \ \ \ \ \ \ \ \ \ \ \ \ \ \ \ \ \
$-\frac{d\mathbb{E}\left[\varPhi\left(0,\theta_1;T\right)%
\right]}{d\theta_1}|_{\theta_1=0}
=\frac{\mu_0\mu_1}{\left(\mu_0-\mu_1\right)\left(\mu_1\right)^2}$

\ 

\ \ \ \ \ \ \ \ \ \ \ \ \ \ \ \ \ \ \ \ \ \ \ \ \ \ \ \ \ \ \ \ \ \ \ \ \ 
$-\frac{\mu_0%
\mu_1\lambda}{\left(\mu_0-\mu_1\right)\left(\lambda+\mu_0-\mu_1\right)\left(%
\mu_1\right)^2}-$
$\frac{\mu_0\mu_1}{\left(\mu_0-\mu_1+\lambda\right)\left(\mu_0+\lambda%
\right)^2},$

\noindent therefore, we have:

\begin{equation}
\mathbb{E}\left[S_1\cdot\boldsymbol{1}_{\left\{S_0<T\leq
S_1\right\}}\right]=\frac{\mu_0}{\mu_1\left(\mu_0-\mu_1+\lambda\right)}
-\frac{\mu_0\mu_1}{\left(\mu_0-\mu_1+\lambda\right)\left(\mu_0+\lambda%
\right)^2}.
\end{equation}

Markovian attack behavior by exploiting the memoryless property of Poisson arrivals has been examined in the single-attack scenario. Marginalization of the joint detection function has produced the probability density of the defense moment together with conditional expectations for the pre-attack and post-attack observation instants. These closed-form results quantify defense timing sensitivity to threat intensity and enable precise calibration of observation parameters for proactive cybersecurity. 

\section{Conclusion} \label{sec04}
This study has presented a stochastic framework for the analysis of cybersecurity defense strategies under a single attack scenario. By modeling the defense process through a continuous observation mechanism with exponential distributions for the defense instant and subsequent slot, the joint detection function has been derived explicitly using Laplace-Carson transforms and first-excess theory. Marginalization under Markovian Poisson attack arrivals has yielded the probability density of the defense moment along with conditional expectations of pre-attack and post-attack observation times. These closed-form results enable quantitative evaluation of defense timing sensitivity to varying threat intensities and facilitate the calibration of observation parameters for enhanced proactive measures.

\backmatter

\bibliography{02_CybA_Reference_R01_amg}
\end{document}